\documentclass[preprint,12pt]{aastex}
\usepackage{natbib,amsmath}
\newcommand{\kms}{km s$^{-1}$}

\newcommand {\apgt} {\ {\raise-.5ex\hbox{$\buildrel>\over\sim$}}\ }
\newcommand {\aplt} {\ {\raise-.5ex\hbox{$\buildrel<\over\sim$}}\ } 
\shorttitle{The Column Temperature}
\shortauthors{Shetty et al.}

\begin{document}
\title{The Effect of Projection on Derived Mass-Size and Linewidth-Size Relationships}
\author{Rahul Shetty\altaffilmark{1,2,5}, David C. Collins\altaffilmark{3}, Jens Kauffmann\altaffilmark{1,2,6}, Alyssa A. Goodman\altaffilmark{1,2}, Erik W. Rosolowsky\altaffilmark{4}, Michael L. Norman\altaffilmark{3}}
\altaffiltext{1}{Harvard-Smithsonian Center for Astrophysics, 60
  Garden Street, Cambridge, MA 02138}
\altaffiltext{2}{Initiative for Innovative Computing, Harvard
University, 60 Oxford Street, Cambridge, MA, 02138}
\altaffiltext{3}{ Laboratory for Computational Astrophysics, Center
for Astrophysics and Space Sciences, University of California
San Diego, LaJolla, CA 92093}
\altaffiltext{4}{University of British Columbia Okanagan, 3333 University Way, Kelowna, BC V1V 1V7, Canada}
\altaffiltext{5}{Current Address: Zentrum f\"ur Astronomie der Universit\"at Heidelberg, Institut f\"ur Theoretische Astrophysik, Albert-Ueberle-Str. 2, 69120 Heidelberg, Germany; rshetty@ita.uni-heidelberg.de}
\altaffiltext{6}{Current Address:  NASA JPL, 4800 Oak Grove Drive, Pasadena, CA 91109}

\begin{abstract}
Power law mass-size and linewidth-size correlations, two of ``Larson's
laws,'' are often studied to assess the dynamical state of clumps
within molecular clouds.  Using the result of a hydrodynamic
simulation of a molecular cloud, we investigate how geometric
projection may affect the derived Larson relationships.  We find that
large scale structures in the column density map have similar masses
and sizes to those in the 3D simulation (PPP).  Smaller scale clumps
in the column density map are measured to be more massive than the PPP
clumps, due to the projection of all emitting gas along lines of
sight.  Further, due to projection effects, structures in a synthetic
spectral observation (PPV) may not necessarily correlate with physical
structures in the simulation.  In considering the turbulent velocities
only, the linewidth-size relationship in the PPV cube is appreciably
different from that measured from the simulation.  Including thermal
pressure in the simulated linewidths imposes a minimum linewidth,
which results in a better agreement in the slopes of the
linewidth-size relationships, though there are still discrepancies in
the offsets, as well as considerable scatter.  Employing commonly used
assumptions in a virial analysis, we find similarities in the computed
virial parameters of the structures in the PPV and PPP cubes.
However, due to the discrepancies in the linewidth- and mass- size
relationships in the PPP and PPV cubes, we caution that applying a
virial analysis to observed clouds may be misleading due to geometric
projection effects.  We speculate that consideration of physical
processes beyond kinetic and gravitational pressure would be required
for accurately assessing whether complex clouds, such as those with
highly filamentary structure, are bound.

\end{abstract}

\keywords{ISM:clouds -- ISM: structure -- methods: analytical -- stars:formation}

\section{Introduction\label{introsec}}

Though stars form in the densest cores within much more voluminous
molecular clouds, the motions and forces within the parent cloud at
various scales significantly shape, if not control, the evolution of
the cores as they form stars.  Observations, in particular of
dust emission and extinction and of a variety of molecular lines, have
provided much information about the internal structure and dynamics of
molecular clouds.  However, determining the 3-dimensional (3D)
structure of the cloud from observations is not trivial, due in large
part to line-of-sight projection effects.

The scaling between the mass $M$ and velocity dispersion $\sigma$ with
size scale is often studied, in both numerical models and observations
of star forming regions \citep[e.g.][Kauffmann et al. 2010a,b in
  preparation]{Ostrikeretal01,
  Myers&Goodman88,Ballesteros-Paredes&MacLow02, Dibetal07,
  Falgaroneetal92,Heyeretal09, Solomonetal87}.  A radius $R$ is often
considered as a proxy for the size of the region under inspection, to
construct power-laws $M \propto R^a$ and $\sigma \propto R^b$.
\citet{Larson81} found $a \sim 2$ and $b \sim 0.5$, now generally
known as (the first and third) ``Larson's Laws."  Larson's second law,
relating $\sigma$ with the ratio of $M$/$R$, is a consequence of the
other two, and is often used to study the dynamic nature of the cloud,
through the virial parameter $\alpha = 5 \sigma^2 R / (MG)$.  The
value of $\alpha$ may be indicative of whether structures or other
such contiguous regions within clouds are bound, due either to its own
self-gravity or by the ambient external pressure
\citep{Bertoldi&McKee92, McKee&Zweibel92}.  However, assumptions about
the virial theorem that are commonly employed to derive $\alpha$,
e.g.\ that the surface terms are negligible compared to the volume
terms, may in fact be erroneous, as discussed by
\citet{Ballesteros-Paredes06} and \citet{Dibetal07}.

In order to properly interpret the Larson scaling relations, a
thorough understanding of the effects of projection would be
necessary.  Contiguous structures in an observed
position-position-velocity (PPV) cube may not be representative of
actual 3D structures in position-position-position (PPP) space of the
simulation\citep{Adler&Roberts92, Ostrikeretal01}.  In fact,
\citet{Pichardoetal00} showed that the structure of a PPV cube is more
tightly correlated with the line-of-sight velocity structure than the
3D density distribution.  Similarly, identified structures in a 2D
(integrated emission and/or extinction) map, such as high density
knots or filaments, may also be a superposition of numerous lower
density peaks along the line of sight
\citep[e.g.][]{Ostrikeretal01,Gammieetal03}.  These projection effects
may indeed provide power law scalings that differ from the actual
scalings \citep[as discussed by][]{Ballesteros-Paredesetal99,
  Ballesteros-Paredes&MacLow02}.  

Here, we assess the effect of projection from an analysis of a 3D
numerical simulation of a molecular cloud.  We compare the derived
$M$-$R$ and $\sigma$ - $R$ relationships from 2D projected density and
3D spectral (PPV) data with those obtained from the full 3D simulation
(PPP) density and velocity data.  To derive $M$, $R$, and $\sigma$, we
employ dendrograms, a recently developed technique which identifies
contiguous structures within various chosen (intensity or density)
thresholds, and in the process characterizes the hierarchical nature
of the data \citep{Rosolowskyetal08}.  We then use the measured sizes,
masses, and linewidths in a virial analysis, to extend our PPP and PPV
comparison.  In the next section, we briefly describe the simulation
dataset and our method of analysis.  In Section 3, we present the
Larson relationships obtained from the full 3D simulation data and
idealized observations of those simulations, and compare the results.
We discuss the implications of the results in Section 4, focusing on
the interpretations of observations.  We summarize our findings in
Section 5.

\section{Method \label{methosec}}

In our investigation of the effect of projection on the derived mass-
size and linewidth-size relationships, we use the result of a 3D
hydrodynamic simulation at a single timestep.  The simulation used for
this study was run with the MHD extension of the Adaptive Mesh
Refinement (AMR) code ENZO described by \citet{Collinsetal09}.  In the
MHD simulation of the molecular cloud, isothermal gas collapses into
filaments and eventually forms dense cores in a 1000 pc$^3$ region,
with periodic boundary conditions.  The gas initially has uniform
density $\rho$ = 200 cm$^{-3}$ and magnetic field $B$ = 0.6 $\mu$G,
with isothermal temperature 10 K.  The virial parameter for this box
is 0.9, giving a slightly unstable initial cloud.  At each timestep,
the gas is driven with a random Gaussian velocity field.  The driving
field has power in a top-hat distribution between wavenumbers $k=1,2$,
and is normalized to keep the energy input constant, as described in
\citet{MacLow99}.  This results in a constant RMS mach number of 9.
This driving was maintained for several dynamical times to obtain
statistical independence from the initial conditions, after which
self-gravity was switched on.  The data analyzed in this study was
taken after 0.5 free fall times.

\begin{figure}
\plotone{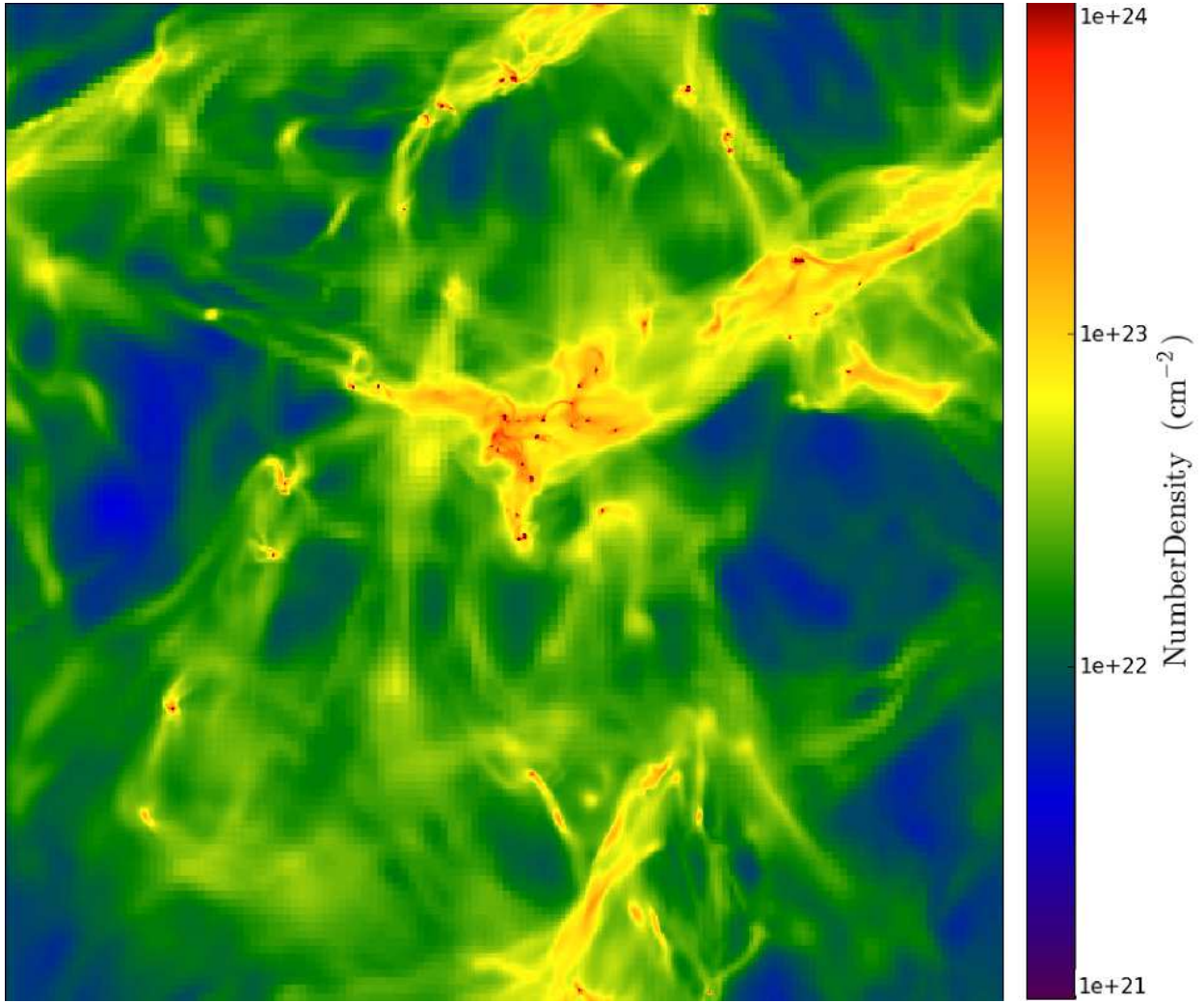}
\caption{Column density of simulated cloud.  Each side has a length of 10 pc.}
\label{simsnap}
\end{figure}

The root grid has a resolution of $128^3$ zones.  Due to the AMR
feature of the ENZO code, the resolution increases with increasing
density, such that the Jeans length of the gas is always resolved by 4
zones, satisfying the Truelove criterion \citep{Trueloveetal97}.  A
total of 4 levels of refinement are added this way.  Self-gravity is
included, by solving the Poisson equation in the root grid using Fast
Fourier Transforms, and in the subgrid patches using a multigrid
technique.  Normalizing the simulation to a 10 pc side length, this
gives a fine grid resolution of $\approx$1000 AU.  A projection of the
density can be seen in Figure \ref{simsnap}, which shows the
filamentary nature of the gas.  Details of this simulation will be
discussed in a forthcoming paper.

A common method to locate clumps involves the identification of
contiguous structures in datacubes above a chosen threshold.  In
algorithms such as {\tt CLUMPFIND} \citep {Williamsetal94}, or similar
variants \citep[see e.g.][]{Dibetal07}, structures are labeled as
clumps if they are distinct from the background or from nearby,
isolated structures.  Some investigators fit Gaussian profiles to
describe the shape of the structures \citep[{\tt
    GAUSSCLUMP},][]{Stutzki&Guesten1990}.  Since molecular clouds are
known to be hierarchical, evidenced by observations of dense knots
situated in filamentary structures within GMCs, such a method may be
inadequate \citep{Pinedaetal09}.  We thus employ ``dendrograms,'' a
technique which characterizes the hierarchical nature of the matter
distribution, while simultaneously identifying contiguous structures
within chosen (intensity or density) thresholds\footnote{We we will
  refer to any of the structures identified by dendrograms generally
  as ``clumps'' regardless of whether they are self-gravitating (or
  bound) or not.}  \citep{Rosolowskyetal08}.

The simulation data provides $\rho$ and the velocity components $v_x$,
$v_y$, and $v_z$ at every position at a chosen time.  The $\rho$-cube
itself contains all the information necessary to obtain the mass and
size distribution of the clumps in the simulation.  The mass of a
clump is simply the density integrated over all zones within a
dendrogram-identified region, or isosurface, in the $\rho$-cube,
multiplied by the volume of each zone.  We regrid the result of the
AMR simulation into a uniform grid with 256$^3$ zones, each with
length $\Delta x$, so that the volume of a zone is $(\Delta x)^3$.  To
characterize the size of each clump, we define a radius $R_{3D}$ as
that which identifies a sphere with the same volume as that bound by
the isosurface, so $R_{3D} \propto \mathcal{N}^{1/3}$, where
$\mathcal{N}$ is the number of zones within the dendrogram defined
isosurface.  We can then assess whether any clear mass-size
relationship exists in the 3D simulation data.

To obtain the linewidth-size relationship of the simulation, we use
the velocity information to measure the velocity dispersion of a
clump.  For any observation, denser gas contribute more to the
observed linewidths than diffuse gas.  Thus, for more direct
comparison with observations, we consider the density weighted
velocity dispersion.  From the isosurface defined in the $\rho$-cube,
the corresponding velocity components $v_x$, $v_y$, and $v_z$, as well
as the density $\rho$, define the 1D density-weighted velocity
dispersion $\sigma_{1D}$ of that particular clump:
\begin{equation}
\sigma_{1D}^2 =
\frac{1}{3}\frac{\sum{\rho[(v_x-\bar{v_x})^2+(v_y-\bar{v_y})^2+(v_z-\bar{v_z})^2]}}{\sum{\rho}},
\label{vdisp}
\end{equation}
where the summation is taken over all $\mathcal{N}$ zones constituting
the identified clump.  Since Equation \ref{vdisp} does not include the
thermal velocity, it is only representative of the non-thermal, or
turbulent, velocities.  An observed linewidth $\sigma_{tot}$ would
include a contribution from the sound speed $c_{s}$ in addition to
$\sigma_{1D}$, so
\begin{equation}
\sigma_{tot}^2 = \sigma_{1D}^2+ c_{s}^2.
\label{vobs}
\end{equation}
In our investigation of the linewidth-size relationship, we consider
both the turbulent linewidth $\sigma_{1D}$ as well as the total
linewidth $\sigma_{tot}$.

In order to investigate the effect of projection, we generate a PPV
cube and a column density map (shown in Figure \ref{simsnap}) of the
simulation cube.  We then produce dendrogram trees of these synthetic
observations, and compare the resulting mass-size and linewidth-size
relationships with those obtained from the full 3D simulation data.
We only consider optically thin emission, so our analysis is analogous
to observational investigations involving much of the volume of the
molecular cloud, including regions containing dense cores
\citep{Myers&Goodman88,Falgaroneetal92,Heyeretal09, Solomonetal87,
  Larson81}.  This study is thus a first step towards a complete
understanding of the unavoidable consequences of projection in
observations.  Results based on this assumption may not be directly
applicable to molecular line observations tracing high densities, such
as ammonia observations of dense star forming cores.  A thorough
investigation of the effect of projection for those high density
tracers would require the additional consideration of radiative
transfer effects.

To produce a 2D column density map, we integrate the density along a
given direction (e.g. $\hat{z}$).  Since we are assuming purely
optically thin emission, the final 2D map is simply the zeroth-moment
of the $\rho$-cube.  We then construct the dendrogram tree of this 2D
map, obtaining the masses and sizes of each 2D-clump.  In this case,
we define the radius $R_{2D} = (\mathcal{N}/\pi)^{1/2}\Delta x$, which
is the radius of a circle with an area identical to the area within
the 2D isosurface.

From the simulation data, a 3D PPV cube is constructed by binning a
chosen velocity component (e.g. $v_z$), and integrating the mass
(e.g. along $\hat{z}$ at each $\hat{x}, \hat{y}$ position) associated
with each velocity bin.  We consider an idealized PPV cube, with high
spatial and spectral resolution of 0.039 pc and 0.025 \kms,
respectively; at these resolutions, both the density and velocity
structures can be assessed to scales smaller than the typical size of
the dense cores within filaments.  Clump masses are obtained by
integrating the intensity within each dendrogram isosurface of the PPV
cubes.  The velocity dispersion is obtained by computing the second
moment of each clump in velocity; we will refer to this moment as
$\sigma_{z}$, since we construct the PPV cube along the $\hat{z}$
direction.  We only consider clouds to by sufficiently resolved if
$\sigma_{z} \ge$ 0.025 \kms, which is the spectral resolution of the
PPV cube \citep{Rosolowsky&Blitz2005}.  For the clumps in the PPV
cube, the observed velocity dispersion $\sigma_{tot}^2 = \sigma_z^2 +
c_s^2$.  As with the 2D map, we use the projected area associated with
each clump to define $R_{2D}$ as the size of the clump.

From the full simulation dataset, we derive the mass-size $M \propto
R_{3D}^a$ and linewidth-size $\sigma_{1D} \propto R_{3D}^b$ and
$\sigma_{tot} \propto R_{3D}^b$ power law relationships of the
simulated cloud.  We then compare those with the $M \propto R_{2D}^a$,
$\sigma_z \propto R_{2D}^b$, and $\sigma_{tot} \propto R_{2D}^b$
relationships obtained from analyses of the column density map and PPV
cube.

\section{Results \label{resultsec}}

\subsection{Mass-Size Relationships}
\begin{figure}
\plotone{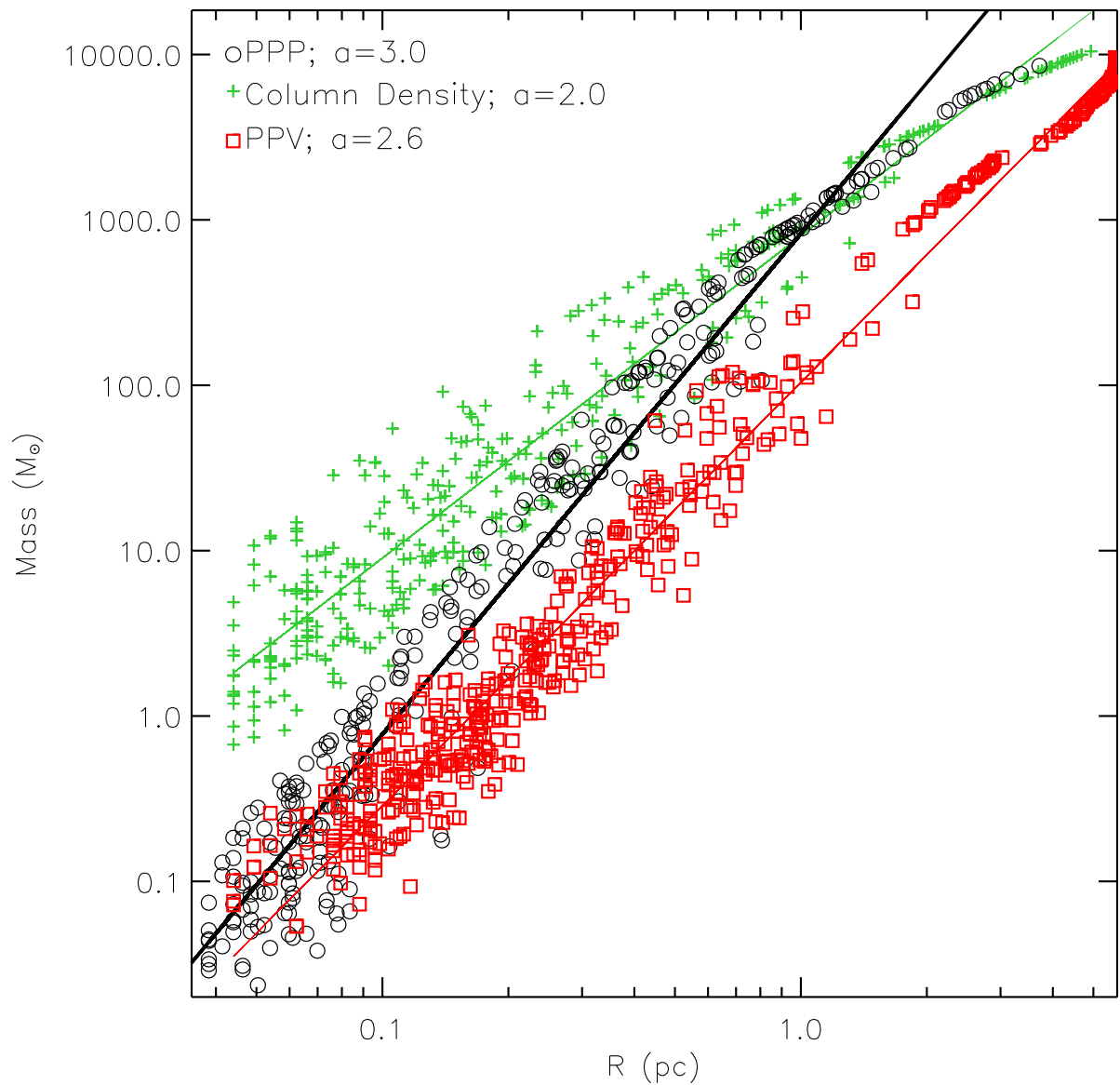}
\caption{Mass-size relationships from dendrogram defined clumps in a
  3D $\rho$-cube (black circles), 2D column density map (green
  crosses), and PPV cube (red squares).  Lines indicate best fits of
  $M \propto R^a$.  In order to distinguish between points, only half
  of the PPP clumps and PPV clumps are shown; the excluded points
  follow the same trends as those shown.}
\label{mrcomp}
\end{figure}

Figure \ref{mrcomp} shows mass-size relationships of the dendrogram
identified clumps, from the full 3D simulation data, the column
density map, and the PPV cube.  In all cases, there is a strong
correlation between the $M$ and $R$, suggestive of a power law
relationship $M \propto R^a$.  Best fit lines give $a \approx 2$ and
$a\approx 3$ for the 2D and 3D cases, respectively, and $a\approx 2.6$
for the PPV cube.  The best fit indices, along with the errors, are
listed in Table \ref{exptab}; the table also indicates best fit
exponents and the errors for the linewidth-size relationships
discussed below.\footnote{The computed (``standard'') errors in the
  linear fits are small due to the very large number of datapoints.
  Thus, the fits provide estimates for the mean value of $M$ or
  $\sigma$ with high accuracy.  However, a prediction of $M$ or
  $\sigma$ using an individual datapoint would have a significant
  error, due to the large scatter in Figures \ref{mrcomp} -
  \ref{lwrth}.}  The best fit mass-size indices from the column
density map and the PPV cube are similar to those derived from many
observations of molecular clouds \citep[e.g. Kauffmann et al. 2010b,
  in preparation,][]{Larson81}.

Power law fits from observations have provided an estimate $a \approx
2$ \citep{Larson81, Solomonetal87}, known as ``Larson's 3rd Law.''
Indices of $a=2$ and $a=3$ indicate that structures have constant
column densities and constant volume densities in 2D and 3D,
respectively.  A further consequence is that the surface density is
constant for all clouds.  

The best fit relationships shown in Figure \ref{mrcomp} suggest that
the dendrogram identified clumps have little density variation within
them.  Clumps with small extents are more likely to have nearly
constant (column or volume) densities, and indeed clumps with $R$\aplt
0.6 pc generally agree well with the $a=2$ and $a=3$ relationships.
Note, however, that at small scales there is still a range of clump
masses at any given $R$, indicating that the fragmentation process
produces clumps with a range of masses.  Larger scale clumps include
contributions from the smaller, high density clumps, and so can have
larger density gradients within their surfaces; as can be seen in
Figure \ref{mrcomp}, those clumps deviate from the $a=2$ and $a=3$
relationships.  We note that we have also modeled the clumps as
ellipsoids \citep[see e.g.][]{Rosolowskyetal08,Bertoldi&McKee92}, and
obtain slightly different mass-size power-law indices, with $a \approx
1.8$ and $a \approx 2.3$ for the 2D map and PPP data, respectively;
this difference indicates that the definition of $R$ plays a role in
the derived mass-size relationship.  Taking these issues into
consideration, even though the best fit indices are $a=2$ and $a=3$
from the column density map and PPP cube, respectively, we do not
conclude that the structures in the simulation have constant volume
densities, or that the surface density is everywhere equivalent.

Though there are strong correlations between $M$ and $R$ at $R$\aplt
0.5 pc, the masses derived from the 2D map are systematically larger
than those from the 3D cube.  As described by \citet{Gammieetal03},
this discrepancy arises because peaks in the 2D map may include
contributions from spatially separated objects which lie along the
same line of sight (see also Kauffmann et al. 2010c, ApJ Submitted,
for mass contamination by extended envelopes); this blending of
structures along the line of sight also results in fewer total clumps
found by dendrograms in the 2D map (not all the clumps are shown in
Figure \ref{mrcomp}).

For clumps with $R$\apgt 0.5 pc, there is relatively good agreement in
the masses of the 2D and 3D structures.  These represent the low
density, large scale structures, and their total masses include the
masses of the higher density, smaller scale clumps embedded within
them (i.e. the mass of the ``branches'' of the dendrogram tree
includes the mass of any ``leaves'' associated with that ``branch,''
see \citet{Rosolowskyetal08} for definitions).  As the clump scale
increases, the masses of both the 2D and the PPP clumps approaches the
total mass of the simulated cloud.

From the PPV cube, many clumps at small scales ($R$\aplt 0.08 pc) have
similar masses and sizes to those from the PPP data.  At those scales,
the clumps are the highest density objects (e.g. ``cores''); many of
those clumps may be detected as objects in high resolution PPV cubes
since they might have velocities that are distinguishable from the
surrounding material.  But, some of the low mass PPV clumps are not
identified as such in the PPP or column density maps; they are simply
part of much larger low density features. They are identified as
clumps in the PPV cube because various regions along that
line-of-sight have similar (turbulent) velocities, and therefore occur
as brighter knots in the PPV cube \citep[as discussed
by][]{Pichardoetal00}.  Thus, many of the low mass, small extent PPV
clumps in Figure \ref{mrcomp} are actually part of the larger, low
density PPP clumps.  Additionally, we found that for PPV cubes with
lower resolution, many of the identified low mass clumps may not be
detected.

At large scales, at a given radius the masses of the PPV clumps are
systematically lower than those of the PPP clumps.  This offset arises
because of the difference in the definition of $R_{2D}$ and $R_{3D}$,
as well as a consequence of only including densities within given
velocity bins in the construct of the PPV cube.  Due to the latter
effect, line of sight velocity gradients within a 3D structure may
result in (1 or more) corresponding structure(s) in the PPV cube
having lower mass(es) than the single 3D object; a 3D structure within
a molecular cloud may thus not appear as a distinct structure in a PPV
cube.  These discrepancies indicate that the identification of clumps
in a PPV cube may not provide accurate estimates of the masses of the
real clumps.

The general agreement between the PPV and PPP masses and sizes at
small scales, transitioning to lower PPV masses at a given size at
larger scales, results in $M\propto R_{2D}^{2.6}$ for the PPV clumps.
A derived index between the $a\approx 2$ result from the 2D analysis
and $a\approx 3$ from the full 3D data should be expected, since a PPV
cube is constructed using the column density in defined line of sight
velocity bins, thus involving a mixture of the column density and 3D
density.

\subsection{Linewidth-Size Relationships}
Figure \ref{lwrcomp} shows the non-thermal $\sigma_{1D}-R_{3D}$ and
$\sigma_{z}-R_{2D}$ relationships of dendrogram identified objects
from the PPP and PPV cubes, respectively.  A best fit of $\sigma
\propto R^b$ produces $b \approx 0.7$ for the PPP case.  For the PPV
clumps, there is a large scatter in the $\sigma_z-R$ relationship, and
a best fit yields $b \approx 0.85$.  In practice, it is difficult to
accurately measure linewidths for regions smaller than a few tenths of
a parsec.  We thus also perform the fit only considering those
structures with $R > 0.2$ pc, and obtain flatter power-laws with $b
\approx 0.5$ for the PPP clumps and $b \approx 0.82$ for the PPV
clumps.  Even when excluding the small scale clumps, a significant
difference in the nonthermal linewidth - size relationship between the
PPP and PPV cases remains.

Besides the differences in the slopes of the linewidth - size
relationships, $\sigma_z$ is systematically lower than the linewidth
computed using all velocity components, $\sigma_{1D}$.  Both PPP and
PPV dispersions are density weighted, either by design (see
Eqn. [\ref{vdisp}]), or due to the intrinsic nature of a PPV cube.
Thus, any discrepancy can be largely attributed to the effect of
projection.  For example, $\sigma_z$ might include contributions from
physically separate structures, since a clump in a PPV cube might
consist of separate structures in the PPP cube.  Additionally,
$\sigma_z$ for a given clump does not include any contribution from
$v_x$ and $v_y$, though it has been estimated that this can account
for at most 20\% of the discrepancy seen.  The observed $\sigma_z$ is
affected by many factors besides the intrinsic velocity distribution
of a given gaseous structure.

As indicated in Section \ref{methosec}, an observed linewidth would
include a contribution from the thermal velocity; Figure \ref{lwrth}
shows the $\sigma_{tot}-R$ relationship from the PPP and PPV data.  At
small scales, the minimum linewidths are 0.2 \kms, which is equal to
$c_s$ of the simulation; turbulence does not contribute much to the
observed linewidths where $c_s >>$ the turbulent velocities.  Compared
with Figure \ref{lwrcomp}, which only shows the turbulent components,
the minimum linewidth imposed by the thermal component forces the
power law index in the PPP and PPV case to decrease to $b=0.44$ and
$b=0.39$, respectively.  For those structures with $R > 0.2$ pc, best
fits do not change the PPP relationship, but increases the PPV
linewidth size index to $b=0.5$, similar to results from numerous
observational works \citep[e.g.][]{Solomonetal87, Larson81}.  Despite
the better correspondence in best fit power laws, there is still a
clear systematic offset between the PPP and PPV total linewidths; the
best fit intercepts still differ by a factor of $\approx$2 (Figure
\ref{lwrth}, with a larger discrepancy in the turbulent offsets, as
evident in Figure \ref{lwrcomp}).  Table \ref{exptab} lists the best
fit indices for the various power laws considered.

\subsection{Mean Subtracted Data and $^{13}$CO Emission}

We note that in a column density map where the mean density was
subtracted off at each location, the resulting mass-size relationship
is similar to that shown in Figure \ref{mrcomp}.  The main difference
is that the measured masses are slightly lower, as would be expected.
The masses of the small scale clumps are still appreciably larger than
those from the PPP cube.  We have also performed our analysis on the
simulation data where only gas above 650 cm$^{-3}$ is considered to be
emitting, a scenario analogous to observing optically thin $^{13}$CO.
We find little difference in the derived mass-size and line-size
relationships compared with the scenario where all gas is emitting.
Again, the main difference is that the measured masses are slightly
lower.

\section{Discussion \label{discsec}}

\subsection{Virial Parameters of the PPP and PPV Clumps}

The observed $M$-$R$ and $\sigma_{tot}$-$R$ correlations have strong
bearings on the interpretation of the state of the cloud, such as the
bounded nature of clumps or the clouds as a whole.  For example, a
relationship between $\sigma$ and $(R/M)$ can be constructed from the
$M \propto R^a$ and $\sigma \propto R^b$ relationships:
\begin{equation}
\sigma \propto (M/R)^\frac{b}{a-1}. 
\label{virrat}
\end{equation}
The relationship expressed by Equation \ref{virrat} is often utilized
for studying whether clumps are bound, through the virial parameter
$\alpha= 5\sigma^2R/(MG)$ \citep[e.g.][]{Goodmanetal09,
  Rosolowskyetal08, Larson81}.  Clumps with $\alpha$\aplt 2 are
considered bound, due to its own self gravity \citep{McKee&Zweibel92}.
For $a=2$ (``Larson's 3rd law'') and $b=1/2$ (``Larson's 1st law''),
$\alpha$ is independent of $R$, and if its value is $\approx 2$, the
clumps under consideration are interpreted to be in, or close to,
virial equilibrium \citep[``Larson's 2nd law,''][]{Larson81}.  We
note, however, that a recent investigation of high resolution
$^{13}$CO observations by \citet{Heyeretal09} has found that
structures in molecular clouds in fact do not universally follow all
of ``Larson laws.''

\begin{figure}
\plotone{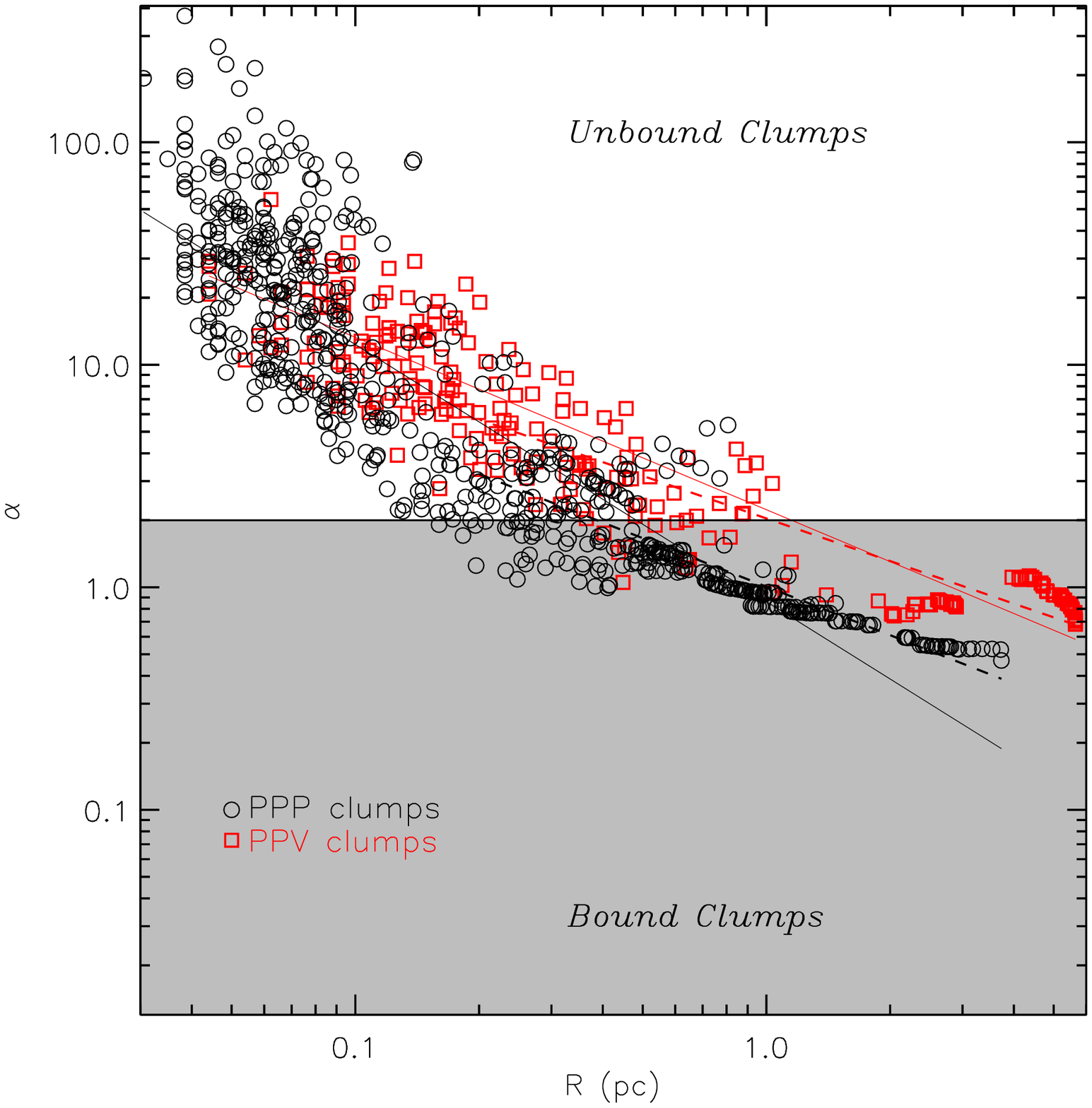}
\caption{Virial parameter ($\alpha$) - size relation for clumps found
  in the 3D simulation (black circles) and synthetic PPV cube (red
  squares).  Best fit lines are also shown, with slopes of $-$1.1 and
  $-$0.8 for the 3D simulation and the PPV clumps, respectively.
  Horizontal line shows $\alpha$=2, indicating virialized clumps.}
\label{virratD}
\end{figure}

Even though we have found significant differences between the power
law relationships we obtain from the PPP and synthetic observations,
we carry forward an analysis to assess the stability of the clumps.
Taking $a=3$ and $b=0.44$ from the PPP cube, $\sigma \propto
(M/R)^{0.22}$.  This relationship leads to $\alpha \propto R^{-1.1}$.
We explicitly show $\alpha$ as a function of $R$ from the PPP analysis
in Figure \ref{virratD}.  Smaller scale clumps have $\alpha$\apgt2,
suggesting they are unbound.  At $R$\apgt 0.5 pc, $\alpha$\aplt2,
suggesting that the large scale structures are close to virialized, or
are bound.

For $a=2.6$ and $b=0.39$ from the PPV analysis, $\sigma \propto
(M/R)^{0.24}$.  These power-laws result in $\alpha \propto R^{-0.8}$.
As Figure \ref{virratD} illustrates, the $\alpha - R$ relationships of
the PPV and PPP clumps are rather similar, despite the glaring
differences in the mass- and linewidth- size relationships shown in
Figures \ref{mrcomp}-\ref{lwrth}.  Though the trend of decreasing
virial parameter with increasing radius from the PPP data is generally
reproduced in the PPV analysis, the threshold radius of $\sim$1 pc
beyond which clouds are bound varies significantly from the
corresponding radius of the PPP clumps.

\begin{figure}
\plotone{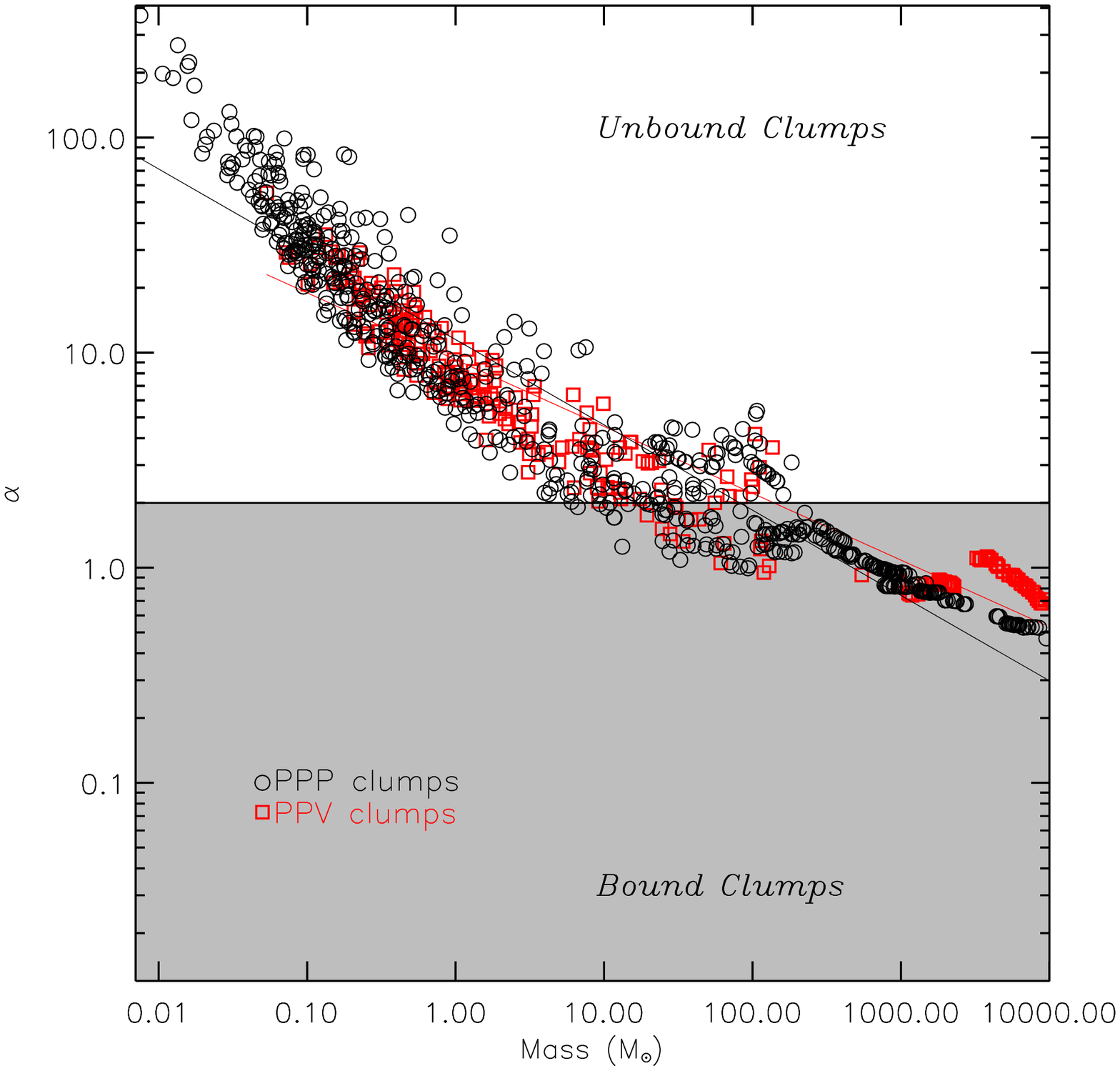}
\caption{Virial parameter ($\alpha$) - mass relation for clumps found
  in the 3D simulation (black circles) and synthetic PPV cube (red
  squares).  Best fit lines are also shown, with slopes of $-$0.4 and
  $-$0.3 for the 3D simulation and the PPV clumps, respectively.
  Horizontal line shows $\alpha$=2, indicating virialized clumps.}
\label{virmass}
\end{figure}

Figure \ref{virmass} shows the $\alpha - M$ relationship.  The slopes
of these power laws are $-$0.4 and $-$0.3 for the PPP and PPV
clumps, respectively.  Similar to the clumps in the simulations of
\citet{Dibetal07}, the large $\alpha$-parameters of the low mass dense
cores suggests that these objects are not bound by their own
self-gravity.  One interpretation of a decreasing virial parameter
with radius, and of very large $\alpha$ for the smallest scale clumps,
is that dense cores are pressure confined, as formulated by
\citet{Bertoldi&McKee92}.  However, our best-fit exponents differ from
the 2/3 value derived for purely pressure confined clumps
\citep{Bertoldi&McKee92}, suggesting that other physical processes,
and/or other terms in the virial equation, need to be taken into
account.

The similarity in the virial parameters of the clumps from the PPP and
PPV clumps must not lead to the interpretation that PPV clumps can
reliably provide accurate measurements of $\alpha$.  One reason for
the general agreement is due to the abundance of low mass clumps in
the PPV cube.  As discussed in $\S$\ref{resultsec}, many of these
clumps in fact are not distinct objects in the simulation data.

Generally, current observations have lower resolutions than those
considered in this work, and such observations would not be capable of
detecting all the small scale PPV clumps shown in Figures \ref{mrcomp}
and \ref{lwrth}.  The resulting $\alpha$-$R$ power law would have a
flatter index than the $-$0.8 shown in Figure \ref{virratD}.  Further,
at intermediate radii (0.1 pc \aplt$R$\aplt 0.3 pc), there is a clear
offset in the measured virial parameter of the PPV clumps compared to
those of the PPP clumps, due to the lower mass estimates of the PPV
clumps (see Figure \ref{mrcomp}).

Additionally, we note that the turbulent linewidths, as opposed to the
total linewidths, produce power law linewidth-size relationships with
markedly greater discrepancy between the PPP and PPV clumps (see
Figure \ref{lwrcomp} and Table \ref{exptab}).  Of course, identifying
the turbulent linewidth is very difficult when the turbulent velocity
components are (very) subsonic; and, the kinetic term in the virial
parameter must include the thermal component to properly assess a
clump's stability.  Yet, the vast differences between the
$\sigma_{1D}-R$ and $\sigma_{z}-R$ relationships are illustrative of
the strong influence of turbulence, in conjunction with projection, on
the measured mass- and linewidth- size relationships.

Nevertheless, $\alpha$ is itself derived by excluding the surface
terms in the virial equation, as well as assuming a negligible
temporal variation in the moment of inertia.  These terms may in fact
be comparable to the surface terms, as
\citet{Ballesteros-Paredesetal99} and \citet{Dibetal07} demonstrated
in extensive analyses of 2D and 3D simulations, respectively.  Other
common assumptions, such as that turbulence only acts against
collapse, may themselves be flawed, as discussed by
\citet{Ballesteros-Paredes06}.  Such simplifications may lead to
inaccurate interpretations of the state of observed clouds.  Given
these caveats, the standard virial analysis may not accurately reveal
the bounded nature of clumps, even when applied to the full 3D
simulation cube.  We thus cannot draw any unequivocal conclusions
about the bounded nature of the clumps in one snapshot of the
simulation.  Our findings simply suggest that commonly assessed
correlations, such as the mass-size and linewidth-size relationships,
may be significantly affected by projection effects.

\subsection{Implications for Interpreting Observations}

The stark discrepancy in the power law relationships between the full
simulation dataset and the synthetic observations may be due in part
to the structure of the simulated molecular cloud (in addition to the
aforementioned choice of the definition of $R$).  In the simulation we
consider, filaments are ubiquitous within the cloud, and most dense
cores are clustered (besides residing in filaments).

We have verified that for purely spherical cores that are completely
isolated (i.e. not lying in the same line-of-sight from other cores)
with distinct velocity profiles, the masses, sizes, and linewidths
derived from the PPP and PPV cubes agree.\footnote{In the simulation,
however, we found that relatively isolated structures give different
best fit power laws from the PPP and synthetic observations.}  If such
``simple'' clouds exists, and given the discrepancy in derived
power-laws between PPP and synthetic observations of the filamentary
simulation we consider, we speculate that there should be some
transition region in parameter space beyond which traditional analysis
methods used to assess the ``boundedness'' of structures cannot be
applicable.  We illustrate this concept in Figure \ref{dangerzone},
which broadly indicates that consideration of more physical processes
is necessary for accurately assessing the boundedness of more complex
clouds.

For a very simple spherical clump, it may be possible to determine if
the object is bound or not using the classical virial parameter
analysis \citep[however, see caveats expounded
  by][]{Ballesteros-Paredes06}.  Including additional physics may
increase the accuracy of the analysis.  It may not be possible to
apply a given analysis, such as the straightforward virial parameter
analysis, to more complex clouds, indicated by the shaded region in
Figure \ref{dangerzone}.  Accurately determining the bounded nature of
objects within complex clouds would require the consideration of more
physics, such as the surface terms in the virial analyses, and/or the
effects of magnetic fields.  

Figure \ref{dangerzone} is only a schematic, intended to illustrate
that considering more physics, rather than just kinetic and
gravitational energies, is required for reliably determining the
nature of the clumps in more complex clouds.  The depiction of a
distinct transition separating the structures for which it is possible
to determine their bounded nature from those for which it is not is
simply an arbitrary illustration.  The confirmation of agreement in
the masses, sizes, and linewidths of simple, isolated cores between
the PPV and PPP cubes is representative of a situation residing near
the origin in Figure \ref{dangerzone} (marked by a circle).  In this
case a simple virial analysis would produce identical results between
a PPP and PPV analysis, and thus may accurately reveal the dynamical
state of the clumps.  On the other hand, our analysis of the highly
filamentary simulation (shown in Figure \ref{simsnap}) clearly lies
within the shaded regime of Figure \ref{dangerzone} where the cloud
structure is rather complex (marked by a cross).  An investigation
into the bounded nature of the clumps in this simulation would require
consideration of more physics than those included in the classic
virial analysis.

The parameter space depicted in Figure \ref{dangerzone} does not
indicate the level of modeling necessary to handle the effect of
projection.  In our analysis, we have simply represented the scale of
the clumps as radii of circles with areas equal to that of the
projected clump.  The simulation we consider is rather filamentary,
and so our method of assigning a ``radius'' to characterize the extent
of the cloud may be partially responsible for the discrepancy in
measured power-law correlations between the PPP and synthetic
observation cases.  As indicated, assuming spherical symmetry may be
sufficient for spherical cores with certain density and velocity
profiles.  However, such idealized cores might not exist in nature,
thus requiring better modeling efforts even for the most simple
objects.

In our analysis, we have not considered effects of chemistry and/or
radiative transfer.  We simply consider a purely optically thin
medium, within which radiation emerges from all matter, or regions
with densities above a threshold density, and assume thermodynamic
equilibrium.  However, the ISM is comprised of various components at
different temperatures \citep[e.g][]{Heiles&Troland03}; individual
cold clouds may also be embedded in warmer gas
\citep{Hennebelle&Inutsuka06}.  The physical state of molecular clouds
may thus be more complex than that considered here.  Further, the
synthetic observations have insignificant noise levels.  Even in
excluding more complex physics and instrumental effects intrinsic to
real observations, we still find rather significant differences
between the measured properties of the cloud from the 3D simulation
data compared with the synthetic observations.  Thus, any
discrepancies in the observed structure, from either a PPV cube or a
column density map, from the 3D structure of the cloud can be fully
attributed to the effect of geometric projection.

Though we have shown that projection may produce inaccurate scaling
relations for a given observed cloud, comparing scaling relations
between various observations may still prove to be worthwhile.  For
example, if analyses of PPV cubes, or 2D column density maps, of
different molecular clouds produce different linewidth- and/or mass-
size relationships, there may be some intrinsic physical process that
could be responsible for the differences (e.g. Kauffmann et al. 2010b,
in preparation).  Some processes, such as heating and cooling, may
play the most influential roles in sculpting one cloud, but may be
insignificant compared to the effect of magnetic fields and gravity in
another; thus, the (observed) scaling relations of those two clouds
could be different.

To infer accurate cloud characteristics from the value of the
exponents of the derived scaling relation, a thorough understanding of
the effect of projection is a necessity.  Analyses of various
simulations could be an avenue toward such an understanding
\citep[][]{Ballesteros-Paredes&MacLow02, Dib&Kim07, Dibetal07}.  In
this work, we have only assessed one particular simulation.  More
analyses on different simulations, e.g. those with different magnetic
field configurations, or including the effects of heating and cooling,
should advance our understanding on {\it how} the ``observed'' mass-
and linewidth- size scaling relations, and ultimately the virial
parameter, varies through the combined effects of geometric projection
and the different physical processes at work.

\section{Summary \label{sumsec}}

We assess the effect of geometric projection in deriving cloud
properties, using a simulation of a molecular cloud.  Using
dendrograms \citep{Rosolowskyetal08}, we identify contiguous
structures in the 3D simulation dataset and idealized synthetic
observations of the simulation.  We measure the masses, sizes, and
linewidths of structures in PPP and PPV cubes, as well as in column
density maps of the simulated cloud.  We subsequently perform a virial
analysis to compare the bounded state of clumps in the simulation with
that assessed from the synthetic observations.  Our main results and
conclusions are:

1) Identified clumps from the 2D column density map with large extents
($R$\apgt 0.8 pc) have masses in agreement with those obtained from 3D
PPP cube.  These large scale structures contain much of the total mass
of the cloud.  However, at smaller scales, the 2D clumps have
systematically higher masses than those from the 3D simulation.  The
measured masses of these smaller scale clumps in the 2D map include
contributions from all gas lying along the same lines of sight,
resulting in inflated mass estimates.

2) Low mass structures with small extents ($R$\aplt0.1 pc) identified
in the PPV cube have similar masses to corresponding objects in the
PPP data.  However, many of these structures are not distinct objects
in the PPP $\rho$-cube; they are identified only because gas from
different regions along (or near) the same line of sight happens to
have similar line of sight velocities.  Further, high spectral and
spatial resolution would be required to identify those structures from
spectral line observations.  On the other hand, at large scales (R
$\apgt$ 0.1), PPV structures systematically have lower masses than PPP
structures.  This discrepancy again arises because of line of sight
effects: a large scale structure in the PPP data might be identified
as numerous lower mass structures in the PPV cube due to gradients in
the line of sight velocity.

\begin{figure}
\plotone{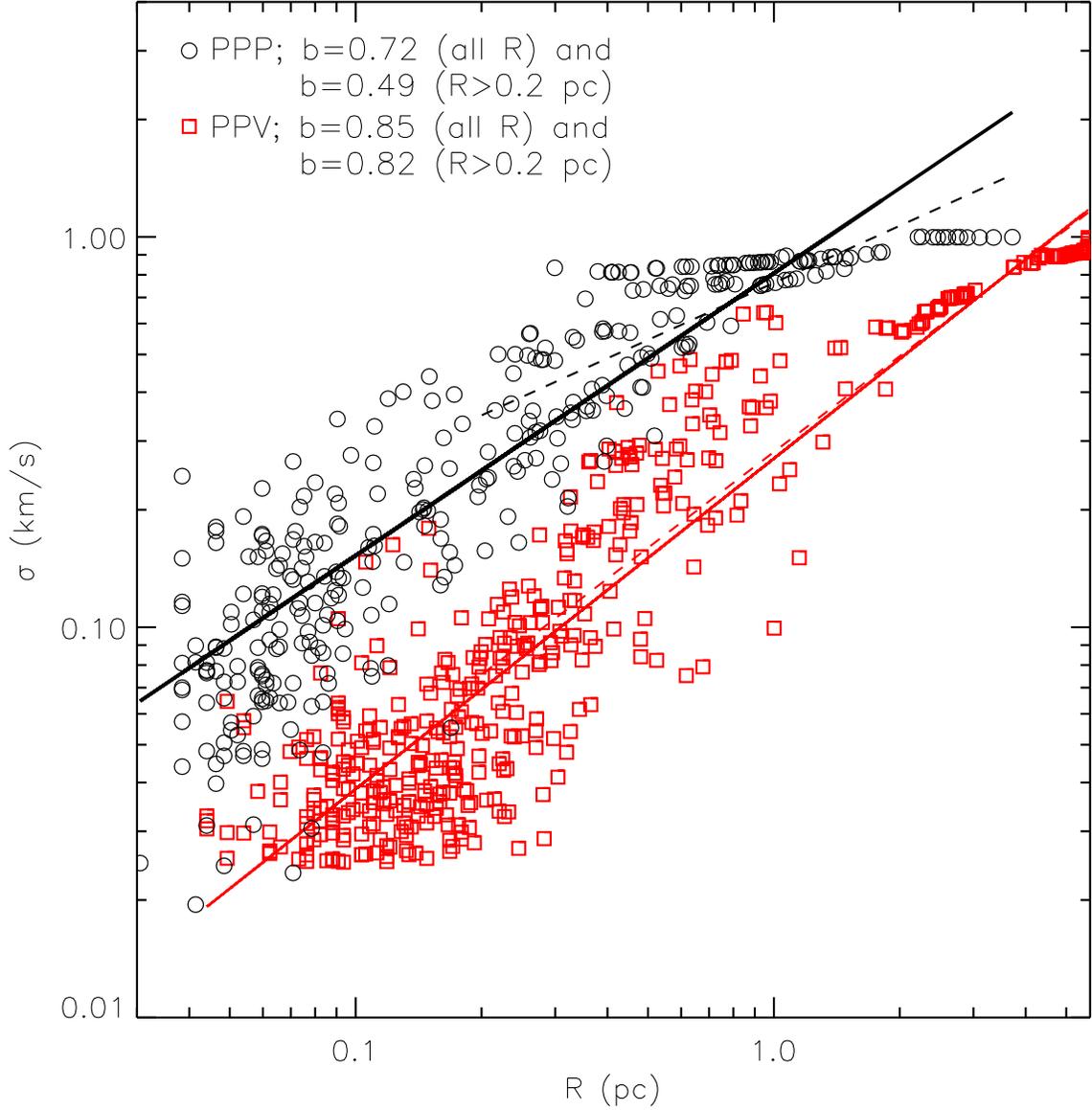}
\caption{Linewidth - size relationship from dendrogram identified
  structures in the PPP $\rho$-cube (black circles) and PPV cube (red
  squares).  Best fit lines of $\sigma_{1D} \propto R^b$ for the PPP
  clumps (black) and $\sigma_z \propto R^b$ for the PPV clumps (red)
  are shown.  Best fits to clumps with $R > 0.2$ pc are also shown
  (dashed lines).  The linewidths ($\sigma_{1D}$ for PPP and
  $\sigma_{z}$ for PPV) do not include the contribution from the sound
  speed.}
\label{lwrcomp}
\end{figure}

\begin{figure}
\plotone{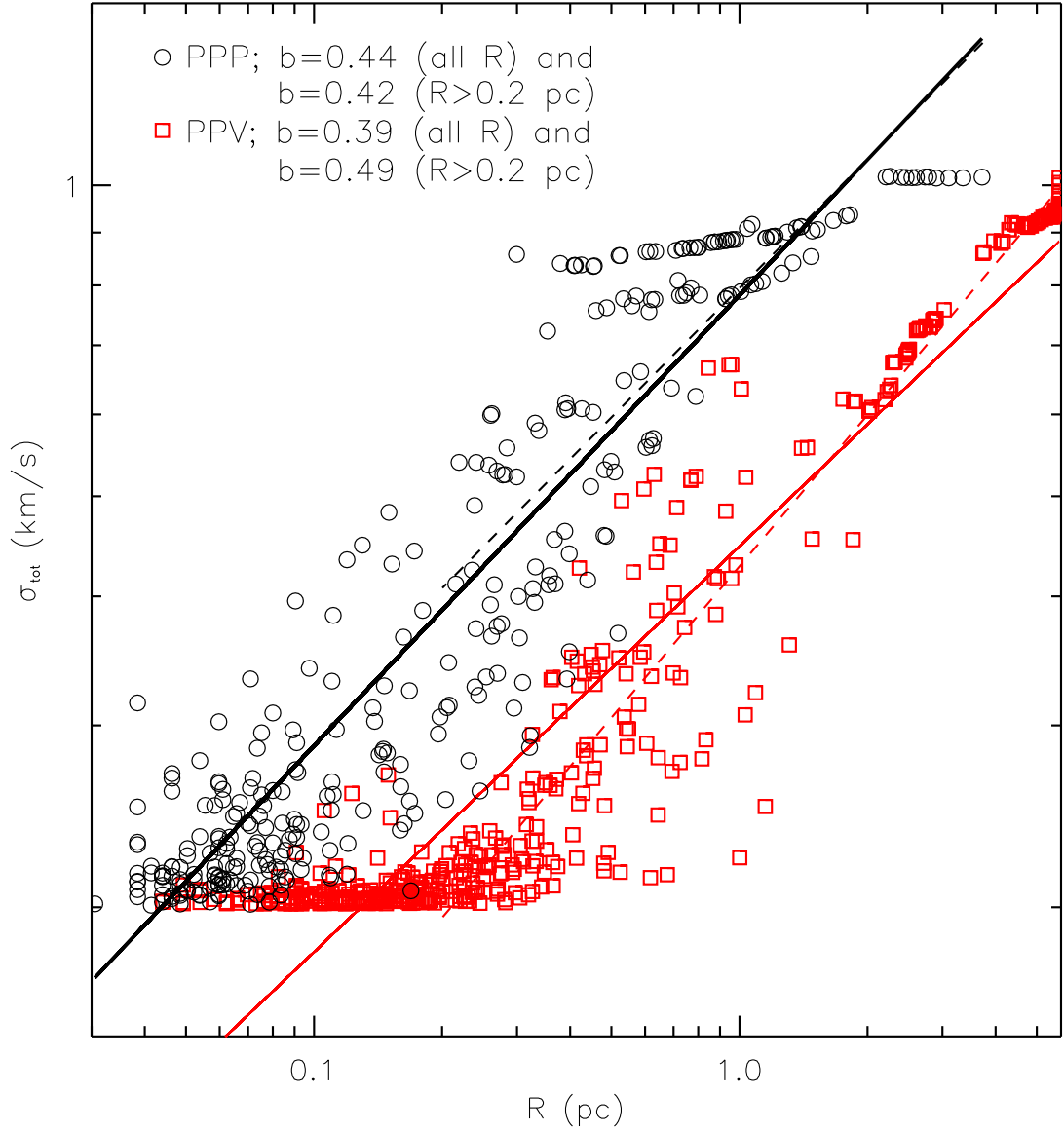}
\caption{Total linewidth - size relationship from dendrogram
  identified structures in the PPP $\rho$-cube (black circles) and PPV
  cube (red squares), along with best fit lines $\sigma_{tot} \propto
  R^b$.  Fits to clumps with $R > 0.2$ pc are also shown (dashed
  line).}
\label{lwrth}
\end{figure}

3) When only turbulent velocities are considered, the cumulative
distribution of clumps from the 3D PPP data give different indices in
the $M$-$R$ and $\sigma-R$ power law relationships compared to those
from the 2D column density map and PPV cube of the simulated cloud.
After including the contribution from thermal pressure, the linewidth
has a lower limit at the value of $c_s$.  This results in similar best
fit $\sigma_{tot}-R$ power-law indices from the PPP and PPV analyses,
though there is a large degree of scatter.  Further, the PPV clumps
systematically have lower linewidths than those of the PPV clumps,
often differing by a factor \apgt 2 (Figures \ref{lwrcomp}-\ref{lwrth}
and Table \ref{exptab}).

4) Due to the differences in the measured properties from the PPP data
and synthetic observations, there is a discrepancy in the identified
scale beyond which the clumps are assessed to be bound.  Despite the
differences indicated by points 2) and 3), a virial analysis of the
clumps in the PPP and PPV cubes show similar trends.  But, we suggest
that the similarity should not lead to the interpretation that a PPV
analysis can accurately reveal the dynamical state of the observed
clumps.

\begin{figure}
\plotone{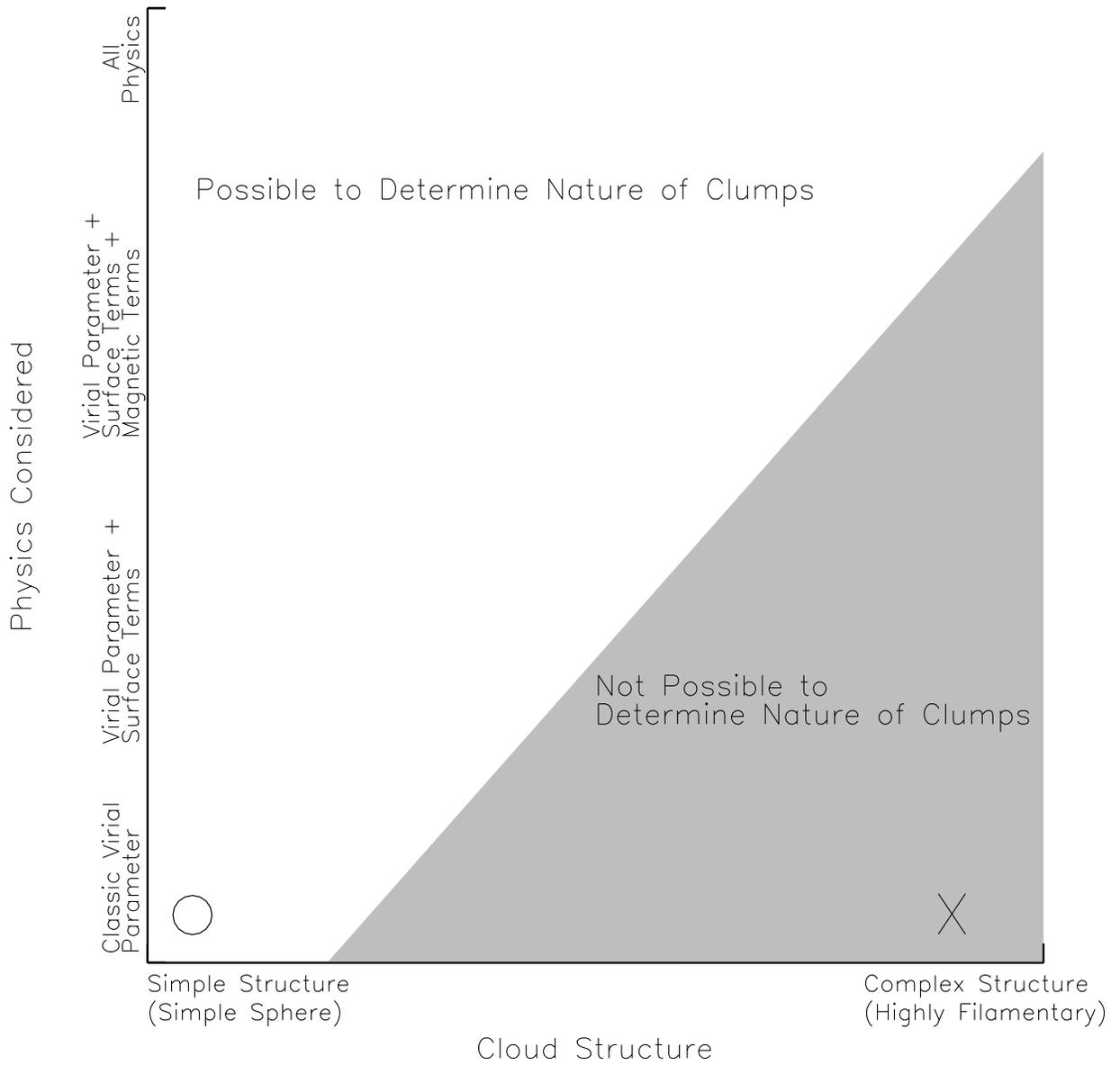}
\caption{Schematic diagram indicating how the consideration of more
  complex physics is possibly required to reliably assess whether
  structures in molecular clouds are bound or not.  The abscissa
  represents the level of complexity in the cloud, from a relatively
  simple sphere to a highly filamentary cloud.  The ordinate
  represents the physical process considered in the analysis.  The
  circle and cross represent cases we have considered in this work.}
\label{dangerzone}
\end{figure}

5) Taking 2) - 4) together, we conclude that projection effects can be
rather significant, leading to inaccurate interpretations of the
dynamical state of the cloud.  We speculate that for a simple
spherical cloud, the classic virial analysis, where the surface and
magnetic terms are omitted (among other assumptions), may be
sufficient for reliably determining whether cores are bound or not.
We suggest that highly filamentary clouds require consideration of
additional physics (Figure \ref{dangerzone}).  We also remark that
better modeling techniques are necessary to properly account for the
effect of projection, as well as to appropriately handle the the
non-spherical shapes of cloud structures.

\acknowledgements

We are grateful to F. Shu for suggesting, after AG's 2008 Tucson
``Stromfest'' talk on the ``self-gravity'' results now found in
\citet{Goodmanetal09}, that we carry out ``tests of the test,'' or, in
other words for suggesting that we investigate the translation and
meaning of the $\alpha$ virial parameter between PPP and PPV space.
We also thank E. Ostriker, P. Myers, and S. Offner for very useful
comments that improved this work.  This presentation benefitted from
many suggestions from an anonymous referee.  We acknowledge use of
NEMO software \citep{Teuben95} to perform our analysis.
D.C. acknowledges support from NSF grant AST0808184, and performed the
simulation at the National Institute for Computational Sciences with
computing time provided by LRAC allocation MCA98N020.  R.S., J. K.,
and A. G. acknowledge support from the Harvard Initiative in
Innovative Computing, which hosts the Star-Formation Taste Tests
Community at which further details on these results can be found and
discussed (see http://www.cfa.harvard.edu/$\sim$agoodman/tastetests).

\bibliography{obsref}

\begin{deluxetable}{ccccc}
  \tablewidth{0pt} \tablecaption{Summary of Power Law Relationships}
\tablehead{ 
\colhead{Power Law; index} & \colhead{PPP\tablenotemark{a}} & \colhead{PPV\tablenotemark{b}} & \colhead{Column Density\tablenotemark{b}}}
\startdata
$M \propto R^a$; $a$ & 3.03 $\pm$ 0.02 & 2.56 $\pm$ 0.01 & 1.95 $\pm$ 0.03  \\

$\sigma_{1D} \propto R^b$; $b$ & 0.72 $\pm$ 0.01 & - & -  \\
$\sigma_{1D} \propto R^b$ ($R>0.2$ pc); $b$ & 0.49 $\pm$ 0.02 & - & -  \\

$\sigma_{z} \propto R^b$; $b$ & - & 0.85 $\pm$ 0.01 & -  \\
$\sigma_{z} \propto R^b$($R>0.2$ pc); $b$ & - & 0.82 $\pm$ 0.01 & -  \\

$\sigma_{tot} \propto R^b$; $b$ & 0.44 $\pm$ 0.01 & 0.39 $\pm$ 0.004 & -  \\
$\sigma_{tot} \propto R^b$($R>0.2$ pc); $b$ & 0.42 $\pm$ 0.02 & 0.49 $\pm$ 0.01 & -  \\
\enddata
{\singlespace 
\tablenotetext{a}{$R=R_{3D}$}
\tablenotetext{b}{$R=R_{2D}$}
}
\label{exptab}
\end{deluxetable}

\end{document}